\documentclass[onecolumn,showpacs,amsmath,amssymb]{revtex4}
\usepackage{enumerate}
\usepackage{dcolumn}
\usepackage{graphicx}
\usepackage{bm}
\begin{document}
\title{Quantization of enlarged Bianchi I universe}

\author{Shintaro Sawayama}
 \email{sawayama0410@gmail.com}
\affiliation{Sawayama Cram School of Physics\\ Atsuhara 328, Fuji-city, Shizuoka prefecture 419-0201, Japan}
\begin{abstract}
We can solve the Wheeler-DeWitt equation of the enlarged Bianchi I as inhomogeneous type universe with static restriction which had been derived by previous paper. 
We consider the commutation relation between the additional constraint and the Hamiltonian constraint.
And we can simplify the usual 3+1 Hamiltonian constraint of the enlarged Bianchi I type universe, and we finally show the state's form is the exponential.
And we find that the states of the quantum gravity are entangled.
Our paper based on the previous work of ours that we call up-to-down method which was aimed to simplify the Wheeler-DeWitt equation.
\end{abstract}

\maketitle
\section{Introduction}\label{sec1}
The quantum gravity is needed because the general relativity is collapsed at two kinds of singularities i.e. initial singularity and black hole singularity.
We treat Bianchi I type universe because of initial singularity and because of resent measurements or study of cosmic microwave backgrounds (C.M.B.).
The enlargement of the Bianchi I type universe has the inhomogeneous information and it may be related to the C.M.B.
Although there are new type of study of quantum gravity which contain loop quantum gravity \cite{As}\cite{Rov}\cite{Thi} or string cosmology \cite{AL}, we use orthodox method of the DeWitt.

Our work based on the method which we call the up-to-down method.
The motivation to think the up-to-down method comes from Isham's quantum category \cite{Isham}.
Based on the Brane world concepts, many Branes are included in the 5-dimensional artificial universe,
and we quantize all the Brane i.e. usual 4-dimensional universe at the same time.
This is the concepts of the up-to-down method.
Our work is based on the motivation of the many world interpretation.
We explain the up-to-down method  \cite{Sa1} in short words.
Once we add an another dimension as external time and secondary we remove the additional dimension, we obtain one additional constraint equation which is related the problem of the time \cite{Sa2}.
The up-to-down method works for two ways.
It works for the way to solve the Wheeler-DeWitt equation \cite{De} and for simplify the mini-superspace \cite{Hart}.
In this paper we calculate the commutation relation between the additional constraint and the Hamiltonian constraint and we find the condition that the both constraints is consistent.
In this paper, $\mu ,\nu$ runs from $0 \to 3$ and $i,j$ runs from $1 \to 3$.

In section \ref{sec2}, we simplify the Hamiltonian constraint.
The Hamiltonian constraint is derived from metric diagonal universe.
In section \ref{sec3}, we search the commutation relation of the Hamiltonian constraint and static restriction.
And we derived the consistent case.
In section \ref{sec4}, we solve the Wheeler-DeWitt equation of the enlarged Bianchi I universe for inhomogeneous case with static restriction.
In section \ref{sec5}, we conclude and discuss the obtained result.

\section{Simplification of the Hamiltonian constraint}\label{sec2}
For simplicity we only treat diagonal metric universes as,
\begin{eqnarray}
\begin{pmatrix}
g_{00} & 0 & 0 & 0 \\
0 & g_{11} & 0 & 0 \\
0 & 0 & g_{22} & 0 \\
0 & 0 & 0 & g_{33}.
\end{pmatrix}
\end{eqnarray}
We start from decomposition of the Einstein Hilbert action of diagonal metric universe as
\begin{eqnarray}
S=\int RdM=\int R[g_{\mu\mu}]dSdt.
\end{eqnarray}
Here $S$ is the hyper-surface with constant time.
Because, our method is different from the usual Wheeler-DeWitt equation formalism,
our obtained Hamiltonian constraint is different type of the Wheeler-DeWitt equation.
If we decompose this action as 3+1, then we can obtain
\begin{eqnarray}
{\cal L}=\dot{q}_{ii}P^{ii}+NH-2\sqrt{q}D^iN_{,i}.
\end{eqnarray}
Here $N$ is the lapse functional and $H$ is the Hamiltonian constraint such that
\begin{eqnarray}
H=\frac{1}{2}q_{ii}q_{jj}P^{ii}P^{jj}+{\cal R}.
\end{eqnarray}
Here ${\cal R}$ is the three dimensional Ricci scalar and $P^{ii}$ is the momentum whose commutation relation with $q_{ii}$ is not $i$, it is $i\sqrt{q}$.
In this formulation there are not appear $q_{ij}$ and $P^{ij}$ and sift vectors and momentum constraint.
So we can ignore the constraint as $[P^{ij},H]$, because we start with metric diagonal setting.
In this simple case we can ignore the diffeomorphism constraints.
If we write the Hamiltonian constraint in the operator representation, we obtain
\begin{eqnarray}
H=\sum_{ij}\frac{1}{2}\frac{\delta^2}{\delta \phi_i\delta \phi_j}+
{\cal R}[q_{11},q_{22},q_{33}]=0 \nonumber \\
=\sum_{ij}\frac{1}{2}\frac{\delta^2}{\delta \phi_i\delta \phi_j}+
\sum_{i\not= j}(\hat{\phi}_{i,jj}+\hat{\phi}_{j,i}\hat{\phi}_{i,i})e^{\hat{\phi}_i}=0.
\end{eqnarray}
If we consider the $\phi$ were only depend $t,x_i$, the Hamiltonian constraint becomes,
\begin{eqnarray}
H\to \sum_{ij}\frac{1}{2}\frac{\delta^2}{\delta \phi_i\delta \phi_j}=0.
\end{eqnarray}
This is the Hamiltonian constraint of the enlarged Bianchi I universe.

\section{Commutation relations between Hamiltonian and additional constraint}\label{sec3}
We treat only diagonal metric case which is $g_{\mu\nu}=0$ for $\mu\not= \nu$.
Then the Hamiltonian constraint ${\cal H}_S$becomes as \cite{Sa3},
\begin{eqnarray}
{\cal H}_S=\sum_{ij}\frac{\delta^2}{\delta \phi_i\delta \phi_j}+\sum_{i\not= j}(\phi_{i,jj}+\phi_{i,i}\phi_{j,i})e^{\phi_i}.
\end{eqnarray}
Here $\phi_i=q_{ii}$ and $g_{\mu\nu}$ means 4-dimensional metric and $q_{ij}$ means 3-dimensional metric and $q_{ii}=e^{\phi_i}$.
The additional constraint ${\cal C} $ of the metric diagonal case is,
\begin{eqnarray}
{\cal C}=\sum_{i\not= j}\frac{\delta^2}{\delta \phi_i \delta\phi_j}+\frac{1}{2}\frac{d}{dt}\sum_i\frac{\delta}{\delta \phi_i}.
\end{eqnarray}
We decompose the additional constraint for static part and dynamical part.
We write the static part as $S$ like,
\begin{eqnarray}
{\cal S}=\sum_{i\not= j}\frac{\delta^2}{\delta \phi_i \delta\phi_j}
\end{eqnarray}
And we also decompose the 3-dimensional Ricci scalar as $\partial \Gamma$ part ${\cal R}^{(1)} $ and $\Gamma \Gamma $ part ${\cal R}^{(2)} $ as,
\begin{eqnarray}
{\cal R}^{(1)}=\sum_{i\not= j}\phi_{i,jj}e^{\phi_i}, \\ 
{\cal R}^{(2)}=\sum_{i\not= j}\phi_{i,i}\phi_{j,i}e^{\phi_i}.
\end{eqnarray}
We calculate the commutation relation of the static restriction and the Hamiltonian constraint by part, then
\begin{eqnarray}
[ {\cal S} ,{\cal R}^{(1)}]=0,
\end{eqnarray}
and
\begin{eqnarray}
[{\cal S},{\cal R}^{(2)}]=\delta^2\sum_{i\not= j}(\phi_{i,ii}+\phi_{i,i}^2+\frac{1}{2} \phi_{i,i}\phi_{j,i})e^{\phi_i}.
\end{eqnarray}
If the right hand side of the above equation is always zero, the static restriction and the Hamiltonian constraint are consistent.
By solving the above equation we obtain one of the $\phi_i$ by other $\phi_j$s.

We define the dynamical part of the additional constraint as ${\cal D}$ as,
\begin{eqnarray}
{\cal D}=\frac{d}{dt}\sum_i\frac{\delta}{\delta \phi_i}.
\end{eqnarray}
Then,
\begin{eqnarray}
[{\cal D},{\cal R}^{(1)}]=\delta \frac{d}{dt} \sum_{i\not= j}\bigg( 2\phi_{i,jj}+\phi_{i,j}^2 \bigg) e^{\phi_i},
\end{eqnarray}
and
\begin{eqnarray}
[{\cal D},{\cal R}^{(2)}]
=\delta \frac{d}{dt}\sum_{i\not= j}\bigg( -\phi_{i,i}\phi_{j,i}-\phi_{j,ii}-\phi_{i,ij}\bigg) e^{\phi_i}.
\end{eqnarray}
Using,
\begin{eqnarray}
\frac{d}{dt}=\sum_i\phi_{i,0}\frac{\delta}{\delta \phi_i},
\end{eqnarray}
we obtain
\begin{eqnarray}
[{\cal D},{\cal R}]
=\delta^2 \sum_{i\not =j}\phi_{i,0}(\phi_{j,ii}-2\phi_{i,ij}-\phi_{i,j}\phi_{i,i})e^{\phi_i}
+\delta^2\sum_{i\not= j}(\phi_{i,0jj}+\phi_{i,0j}\phi_{j,j}-\phi_{i,0ji}-\phi_{i,0i}\phi_{i,j})e^{\phi_i} 
\end{eqnarray}
Then the commutation relation of the additional constraint and the Hamiltonian constraint becomes as,
\begin{eqnarray}
[{\cal C},{\cal R}]=\delta^2\sum_{i\not= j}(\phi_{i,ii}+\phi_{i,i}^2+\frac{1}{2} \phi_{i,i}\phi_{j,i})e^{\phi_i} 
-\frac{1}{2}\delta^2 \sum_{i\not =j}\phi_{i,0}(\phi_{j,ii}-2\phi_{i,ij}-\phi_{i,j}\phi_{i,i})e^{\phi_i} \nonumber \\
-\frac{1}{2}\delta^2\sum_{i\not= j}(\phi_{i,0jj}+\phi_{i,0j}\phi_{j,j}-\phi_{i,0ji}-\phi_{i,0i}\phi_{i,j})e^{\phi_i} .
\end{eqnarray}
This is the commutation relation of the additional constraint and the Hamiltonian constraint.

\section{Solving the Wheeler-DeWitt equation}\label{sec4}
There exist two kind of universe whose commutation relation is always zero which is,
\begin{eqnarray}
\begin{pmatrix}
g_{00} & 0 & 0 & 0 \\
0 & q_{11}(x_1) & 0 & 0 \\
0 & 0 & q_{22}(x_2) & 0 \\
0 & 0 & 0 & q_{33}(x_3) 
\end{pmatrix}\label{metric}
\end{eqnarray}
and
\begin{eqnarray}
\begin{pmatrix}
g_{00} & 0 & 0 & 0 \\
0 & g_{11}(x_2,x_3) & 0 & 0 \\
0 & 0 & g_{22}(x_1,x_3) & 0 \\
0 & 0 & 0 & g_{33}(x_1,x_2)
\end{pmatrix}.
\end{eqnarray}
In this two case we can ignore the all the commutation relation.
We only consider first case.

If the metrix became as Eq.(\ref{metric}) with cosmological constant, the Hamiltonian constraint becomes as
\begin{eqnarray}
{\cal H}_S=\sum_{ij}\frac{\delta^2}{\delta\phi_i\delta \phi_j}-\Lambda =0 .\label{f1}
\end{eqnarray}
And the static restriction with the cosmological constant becomes as
\begin{eqnarray}
{\cal S}=\sum_{i\not= j}\frac{\delta^2}{\delta\phi_i\delta\phi_j}-\Lambda=0
\end{eqnarray}
The solution which satisfy both constraints equation i.e. Hamiltonian constraint equation and the static restriction has form as
\begin{eqnarray}
|\Psi^4\rangle =\prod_i f_i[\phi_i].
\end{eqnarray}
The solution which satisfy both constraint has the form of
\begin{eqnarray}
|\Psi^4(\phi)\rangle =\prod_i \exp (a_i\Lambda^{1/2}\int \delta \phi_i (x_i))  \\
=\prod_i\exp (a_i\Lambda^{1/2}\int \phi_i (x_i)d\Sigma )
\end{eqnarray}
This is the main result of ours.
Here, $d\Sigma$ means integral of on the hypersurface.
And here, $a_1,a_2,a_3$ is the constant factor which satisfy
\begin{eqnarray}
a_1a_2+a_1a_3+a_2a_3=1 \\
a_1^2+a_2^2+a_3^2=1.
\end{eqnarray}
Because there are three parameter and two equation for $a_i$,
the solution of the $a_i$ are continuously infinite.
So the basis of the state is infinite.
If one of the norm of $a_i$ should be one we obtain
\begin{eqnarray}
a_1=\pm \sqrt{2}, a_2=\pm i,a_3=\mp i\\
a_1=\pm i, a_2=\pm\sqrt{2}, a_3=\mp i \\
a_1=\pm i, a_2=\mp i , a_3=\pm\sqrt{2}.
\end{eqnarray} 
We can easily find above state is entangled.

And we should enlarge this state to the $|\Psi^{5(4)}(\phi_{\mu})\rangle$ state, because we use the up-to-down method.
We should enlarge it by the equation of
\begin{eqnarray}
m\hat{H}_S=\sum_{\mu\not= \nu}\frac{\delta^2}{\delta\phi_{\mu}\delta\phi_{\nu}}-\frac{1}{2}\dot{P}=\Lambda .
\end{eqnarray}
If we assume $g_{00}$ does not depend on time $t$, the $\dot{P}$ term vanishes.
And we can easily find the enlargement by assuming
\begin{eqnarray}
|\Psi^{5(4)}(\phi_{\mu})\rangle =f[\phi_0]|\Psi^4(\phi_i)\rangle .
\end{eqnarray}
Then $f[\phi_0]$ has equation of 
\begin{eqnarray}
(a_1+a_2+a_3)\frac{\delta}{\delta \phi_0}f[\phi_0]=0 .
\end{eqnarray}
So the $f[\phi_0]$ is constant, and then $|\Psi^{5(4)}(\phi_{\mu})\rangle$ and $|\Psi^4(\phi_i)$ is same.
However, the functional space is different.
And we can know the projection of ${\cal H}_{5(4)}$ to the ${\cal H}_{4(5)}$ has non zero meager.
So the up-to-down method can be applied.

\section{Conclusion and discussions}\label{sec5}
We analyzed the commutation relation of the Hamiltonian constraint of the metric diagonal case and static restriction of the quantum gravity. 
And we found the case that the Hamiltonian constraint and static restriction is always commute.

We quantized static and inhomogeneous Bianchi I universe.
The result is the summation of the cosine functional.
However, the result is different for homogeneous case.
In the inhomogeneous case infinite product appear.
The freedom of parameter is continuous infinite and the number of freedom is the number of function at each point.
The result is similar to the path integrals. 

From the obtain result, we can say the state of the quantum gravity are entangled \cite{NC}.
The entanglement is the one of the main concepts of the quantum information theory.
A.Hosoya predict the state may be entangled \cite{Ho}.
Because x,y and z direction are entangled.
If we measure x direction we can know information of y and z direction.
And if we used field instead of cosmological constant, we can know philosophical question what is existence by physics.

\end{document}